\documentclass[pre,a4paper,twocolumn,floatfix]{revtex4-1}
\usepackage[utf8]{inputenc}
\usepackage[T1]{fontenc}
\usepackage{lmodern}
\usepackage{amssymb}
\usepackage{amsmath}
\usepackage{amsbsy}
\usepackage{bm}
\usepackage{esint}
\usepackage{braket}
\usepackage{enumitem}
\usepackage{graphicx}
\usepackage{color}
\usepackage{url}
\usepackage{siunitx}
\usepackage{hyperref}
\hypersetup{
  colorlinks=true,
  linkcolor=blue,
  filecolor=blue,      
  citecolor=blue,
  urlcolor=blue,
}
\newcommand{\I}{\mathrm{i}}

\DeclareMathOperator{\Tr}{Tr}
\DeclareMathOperator{\len}{len}

\begin{document}
\title{Thermal state entanglement entropy on a quantum graph}
\author{Alberto D.\ Verga}\email{alberto.verga@univ-amu.fr}
\affiliation{Aix-Marseille Université, CPT, Campus de Luminy, case 907, 13288 Marseille, France}
\author{Ricardo Gabriel Elías}%
\affiliation{Departamento de Física and CEDENNA, Universidad de Santiago de Chile, Avda.\ Ecuador 3493, Santiago Chile\email{gabriel.elias@usach.cl}}
\date{\today}
\begin{abstract}
  A particle jumps between the nodes of a graph interacting with local spins. We show that the entanglement entropy of the particle with the spin network is related to the length of the minimum cycle basis. The structure of the thermal state is reminiscent to the string-net of spin liquids.
\end{abstract}
\maketitle

% text
\section{Introduction}

In 1935 Schrödinger defined entanglement as the distinctive property of quantum mechanics \cite{Schrodinger-1935}. It is interesting to note that, in addition to the usual definition in terms of the separability of a composite system state (the ``representative'' in his words) into a product of individual factors corresponding to each constituent, he put forward the fact that the information contained in the whole cannot be necessarily obtained from the information contained in its parts, implicitly expressing that entanglement is an information resource. Since then the double status of the quantum state as physical and informational has gained in importance to characterize a great variety of phenomena \cite{Stanescu-2016,Zeng-2019}, ranging from quantum computing \cite{Nielsen-2010fk,Georgescu-2014} to topological phases in condensed matter \cite{Fradkin-2013,Wen-2017}.

Highly entangled states are the essential resource used in the so called measurement based quantum computing \cite{Raussendorf-2003rm}, in which one qubit measurements are sequentially realized on a previously build cluster state, a quantum state associated to a graph. The structure of entangled states is also important in the characterization of topological phases \cite{Laflorencie-2016fk}, where the long range entanglement is related to the string-net organization of the gapped ground state \cite{Wen-1990ty,Levin-2005}. Another important domain where entanglement plays a central role is in the relaxation of an initially out of equilibrium state, towards thermal equilibrium \cite{Borgonovi-2016qe,Alessio-2016fj}. As stated by the eigenvalue thermalization hypothesis, a necessary condition for an isolated system in a pure state to reach a thermodynamic state (described by the microcanonical ensemble), is that the energy eigenvectors are chaotic \cite{Deutsch-1991vn,Srednicki-1994ys,Rigol-2008uq}. 

Entanglement is not only a theoretical concept, but it is also a measurable quantity. The von Neumann and Rény entropies are used, among other indicators, to quantify entanglement \cite{Horodecki-2009yg}; these quantifiers can be experimentally accessed, for instance, with cold atoms in optical lattices \cite{Islam-2015,Lukin-2019}. Spatially resolved entanglement entropy measurements can be useful to detect the nonlocal, topological, order in quantum spin liquids \cite{Balents-2010fk,Jiang-2012}. In a different context, the relaxation of isolated systems towards equilibrium, measurement of entanglement also shed light on the underlying mechanism: entanglement entropy of a subsystem identifies with the thermal entropy \cite{Kaufman-2016mz} (see also Ref.~\cite{Neill-2016fr}). Thermal states of isolated systems and spin liquid ground states are examples of highly entangled states, and although fundamentally different, they share structural features as revealed for example by the finite temperature behavior of the Kitaev hexagonal model \cite{Rousochatzakis-2019,Kitaev-2006dn}.

A paradigmatic example of string-net ground state is given by the toric code of Kitaev \cite{Kitaev-2003fk}. The toric Hamiltonian is constructed in such a way that its ground state is invariant under a set of local operators, in analogy with stabilizer codes used in error correction algorithms \cite{Terhal-2015}. Kitaev showed that the protected subspace of the toric code, the stabilizer subspace, coincides with the ground state of the toric Hamiltonian \cite{Kitaev-2003fk}. The toric model can be worked out exactly, and, as first found by Hamma et al.\ \cite{Hamma-2005}, the entanglement entropy reveals the string-net structure of the ground state \cite{Levin-2005}. The large ground state entropy and strong quantum fluctuations inherent to the topological phase of frustrated systems, allow to recover the low-temperature properties of quantum liquids using a single many-body eigenstate \cite{Rousochatzakis-2019}. This remarkable universality of the many-body eigenstates was called typicality in Lloyd thesis \cite{Lloyd-2013}, and in fact it can be seen as a consequence of the eigenvalue thermalization hypothesis \cite{Rigol-2012yq,Mori-2018}. Therefore, in spite of their seemingly different physics, a link can be established between some quantum liquids and thermalized systems through the complex structure of their many-body, highly entangled, quantum states.

In a previous work \cite{Verga-2019} we investigated the entanglement properties of graph quantum states, obtained by the interaction of a particle walking between neighboring nodes and the network of spins. The model can be viewed as a unitary construction of a quantum state associated to a graph by the interaction of a walker with the spins (which physically define the graph connectivity). As such, the system generalizes the simple quantum walk \cite{Kempe-2003fk,Reitzner-2011dq}, used in particular to study topological phases in condensed mater \cite{Kitagawa-2012fk,Simon-2018}, and quantum states associated to graphs, used for instance, to investigate the graph structure, or more generally, multipartite quantum states as a computational resource \cite{Markham-2008,Ionicioiu-2012kx,Garnerone-2012uq,Perseguers-2013uq,Biamonte-2019}. 

We found that, for rather general graphs and a set of simple rules for the motion and interactions, the particle-spins system evolves towards a thermal stationary state well described by the eigenvectors of the unitary operator. The purpose of this paper is to further investigate the entanglement structure of the obtained random thermal state. More specifically, we analyze the entanglement between the walker and the spin network, and propose an explicit formula for the von Neumann entropy, based on the assumption that the structure of the thermal state is somewhat similar to the ground state of spin liquids (string-net): it can be described as a superposition of closed strings, or more precisely, as an independent set of cycles belonging to the cycle graph space.

In the next section we recall the main ingredients of the model, Hilbert space and unitary operators associated with the graph. Next, we formulate our hypothesis concerning the structure of the thermal state, and propose a formula of the particle-spin entanglement entropy. To assert this formula we compare its prediction with a series of numerical results on random graphs. The last section contains our conclusions.

\begin{figure}
  \centering
  \includegraphics[width=0.33\textwidth]{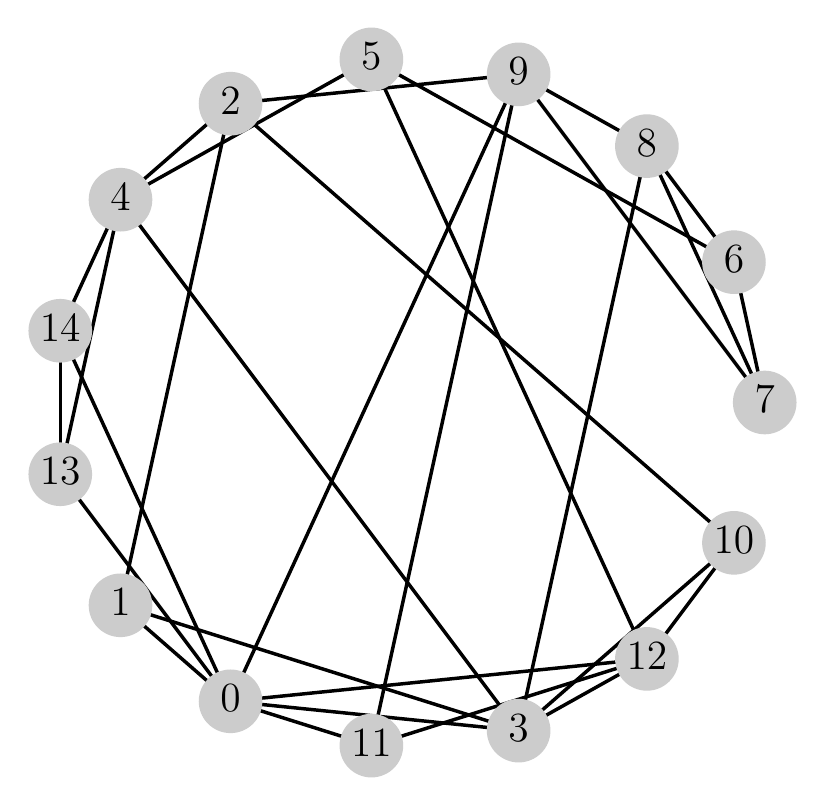}\\
  \includegraphics[width=0.33\textwidth]{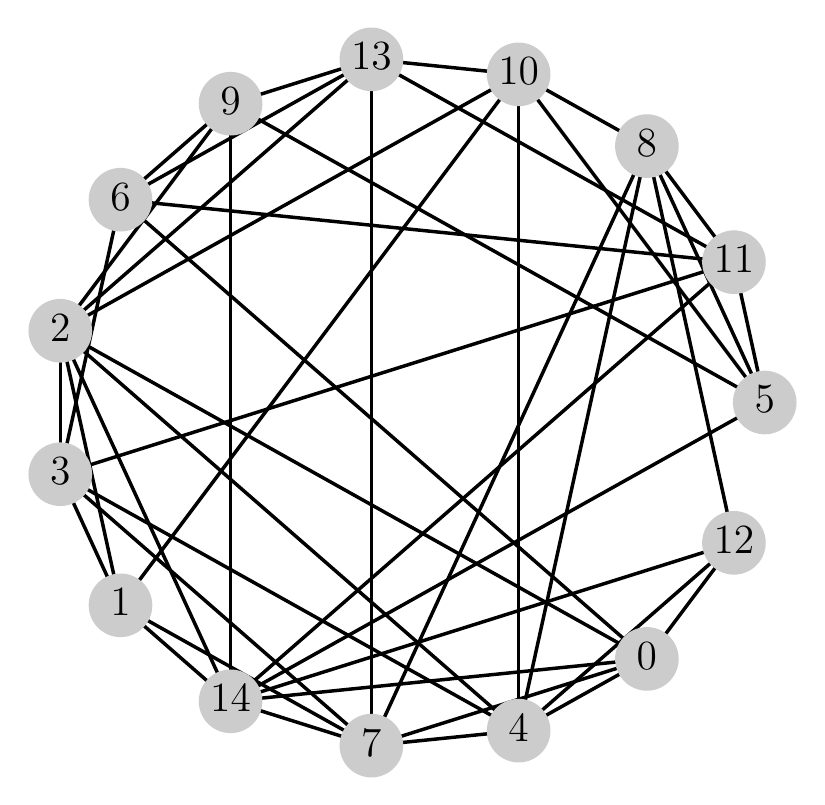}
  \caption{Typical random graphs: (top) Watts-Strogatz with $15$ vertices, an average of $4$ neighbors and rewiring probability $0.35$, (bottom) Erdős-Rényi with $15$ vertices, mean degree $6$, and edge creation probability $0.35$. The corresponding Hilbert space dimensions are \num{1966080} and \num{2949120}.
  \label{f:g}}
\end{figure}

\section{Quantum walk on a spin network}
\label{s:model}

\begin{figure*}
  \centering
  \includegraphics[width=0.33\textwidth]{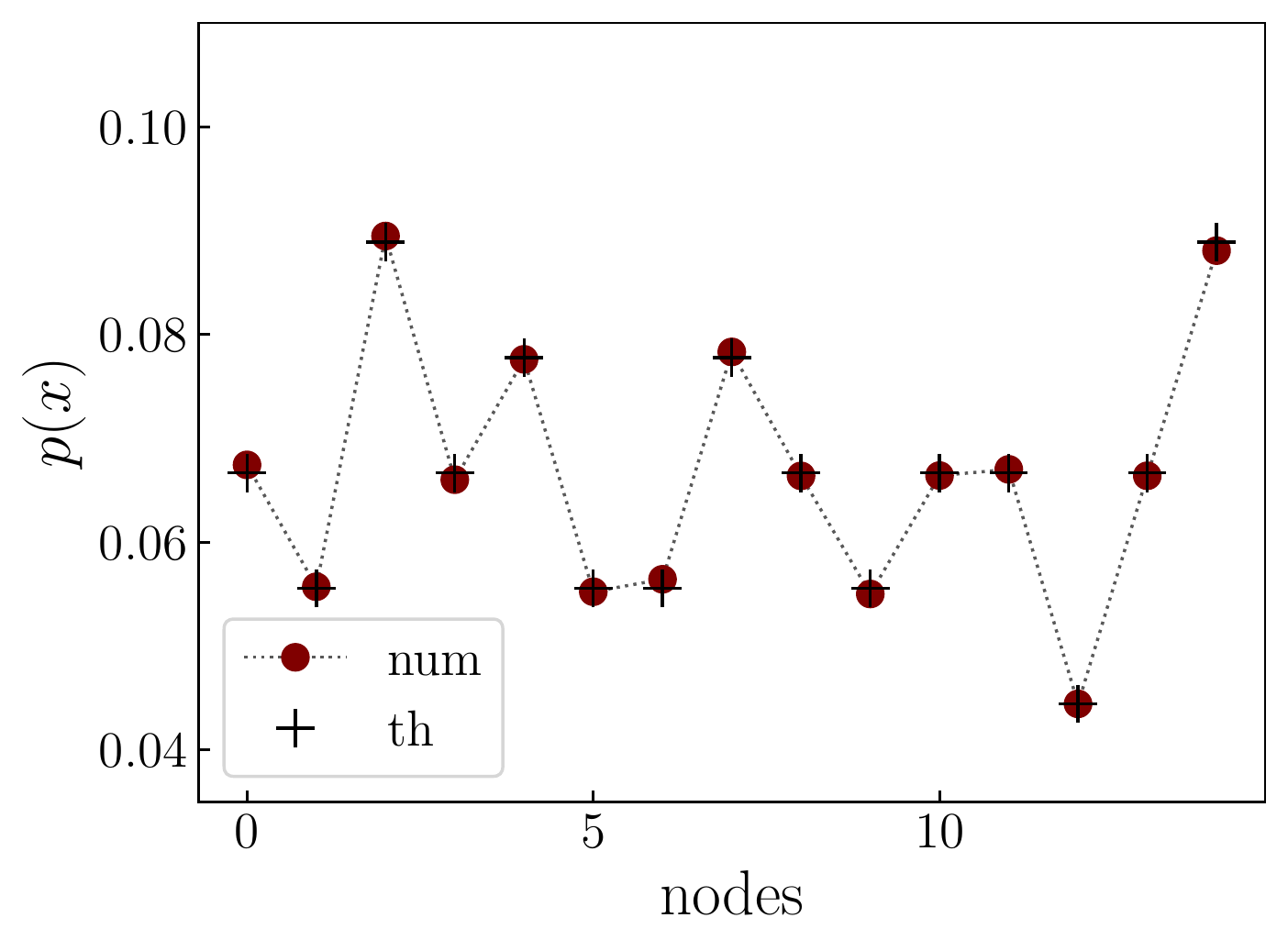}%
  \includegraphics[width=0.33\textwidth]{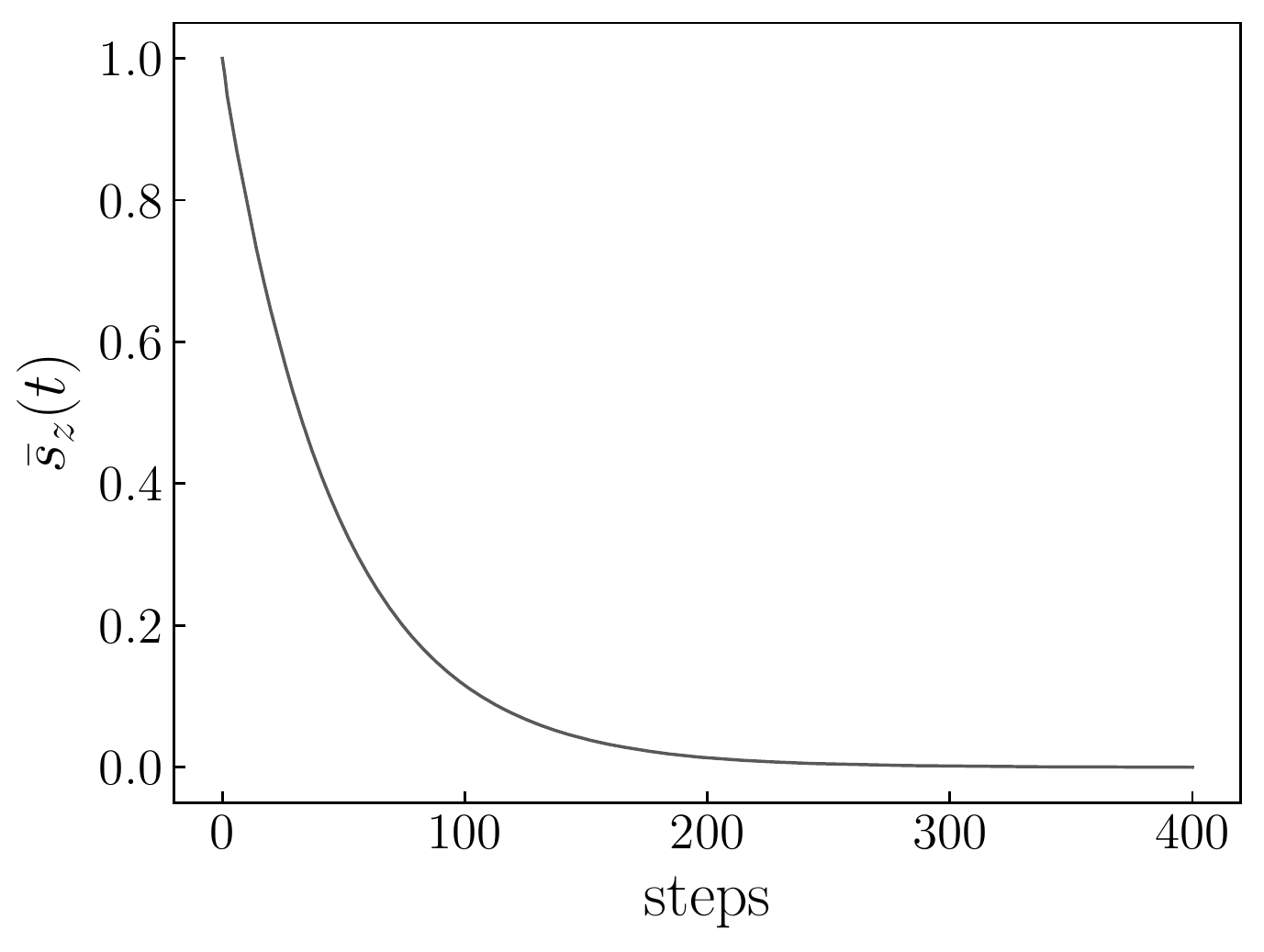}%
  \includegraphics[width=0.33\textwidth]{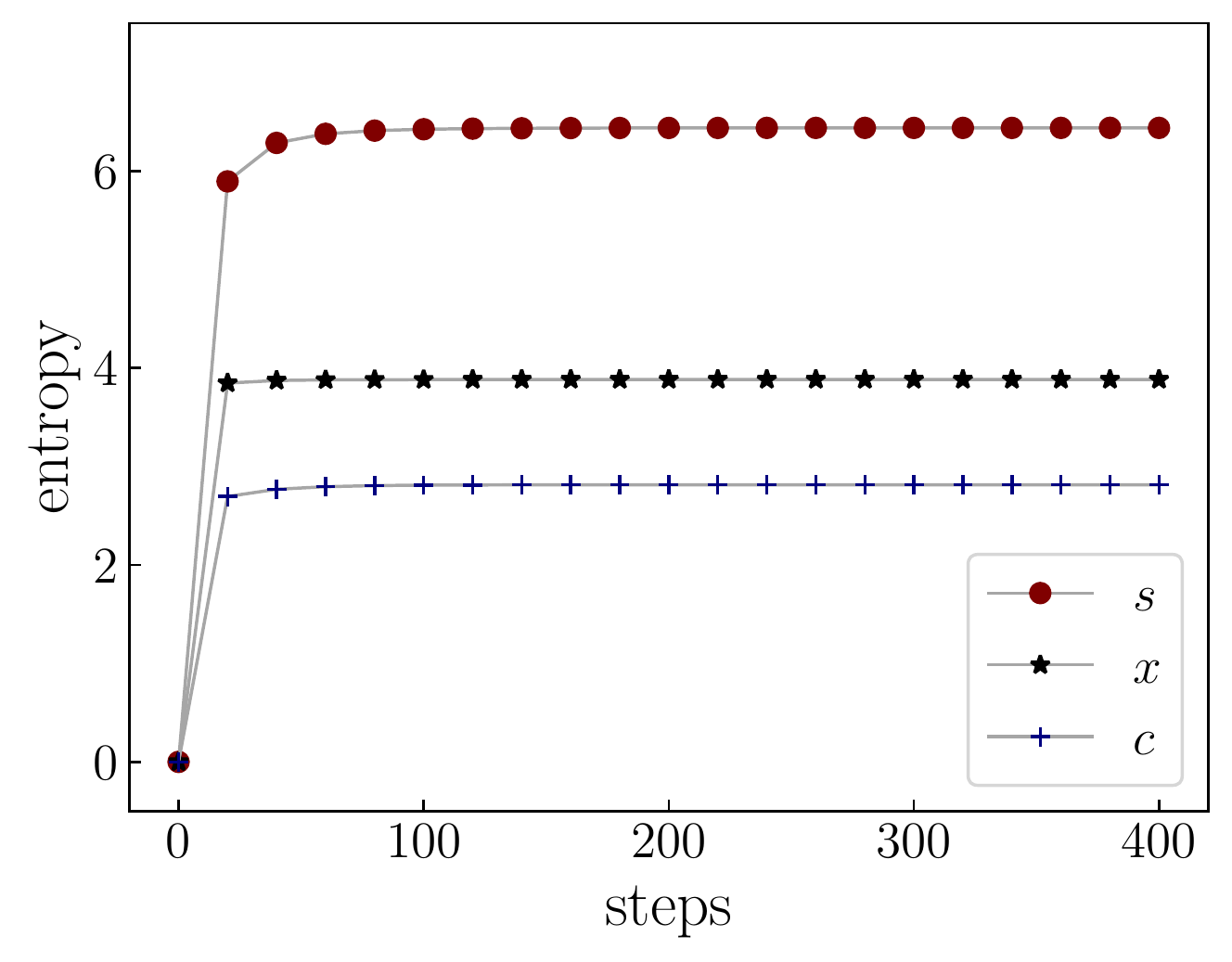}
  \caption{Thermal properties of the stationary state. (left) Position probability (`num' dotted line) showing good correspondence with the microcanonical distribution (`th' crosses). (center) Mean spin $z$ component as a function of time, showing a relaxation to the paramagnetic state. (right) Evolution of the entanglement entropy for the position \(x\), color \(c\), and spin \(s\) (ER random graph with 15 nodes).
  \label{f:TH}}
\end{figure*}

We consider quantum systems \cite{Verga-2019} defined on graphs \(G = (V, E)\), consisting in a set \(x \in V\) of \(|V|\) vertices (\(|\cdot|\) denotes the cardinal of a set), and a set \(e \in E\) of \(|E|\) edges; the set of neighbors of the node \(x\) is,
\begin{equation}
  \label{e:E}
  V_x = \{ y \,|\, e = (x,y) \in E \}\,,
\end{equation}
whose size is \(d_x = |V_x|\), the degree of node \(x\). A particle, the walker, can jump between neighboring nodes of \(G\); its state is characterized by the position quantum number \(x\), referring to the node, and an internal degree of freedom, the color \(c\), referring to its neighbors; they take values in the sets \(x = \{0, \ldots, |V| - 1\}\) and \(c = \{0, \ldots, d_x-1\}\). In addition, on each node sits a one half spin labeled by \(s_x = 0, 1\), \(x \in V\), for the spin up and down states, respectively. The spin network and the particle define together a Hilbert space \(\mathcal{H}\), spanned by the basis
\begin{equation}
  \label{e:xcs}
  \ket{xcs} = \ket{x}\otimes \ket{c} \otimes \ket{s_0s_1\ldots s_{2^{|V|}-1}} \in \mathcal{H}\,,
\end{equation}
where \(s = s_0s_1\ldots s_{|V|}\) is a \(|V|\)-digits binary string representing the \(2^{|V|} - 1\) spins states. We choose two different types of random graphs, the Watts–Strogatz (small world) \cite{Watts-1998rc} and Erdős-Rényi graphs \cite{Erdos-1959fk}, having \(|V| = 6, \ldots, 15\) nodes. The Watts-Strogatz graph (WS), is chosen to have a mean of $4$ neighbors per node and rewiring probability of \num{0.35}. For the Erdős-Rényi graph (ER), the probability to create an edge is fixed to \num{0.35}, which gives, according to the increasing number of vertices, mean degrees between 3 and 6, approximately. The Hilbert space dimension of the systems corresponding to the largest graphs is about \num{3E6}. Figure~\ref{f:g} presents an example of two 15 nodes graphs used in the computations. The graphs were generated with the \texttt{networkx} python library.

One step of the quantum walk evolution operator \(U\) decomposes into three parts,
\begin{itemize}
  \item an operation over the color degrees of freedom (Fourier coin \cite{Shor-1994qr,Kempe-2003fk})
    \begin{equation}
      \label{e:Cop}
      C(d_x) = \frac{1}{\sqrt{d_x}} \exp \left(2\I \pi  \bm c \wedge \bm c/d_x\right)\,,
    \end{equation}
    where \(\wedge\) denotes the outer product, and \(\bm c = (0,1,\ldots,d_x)\) is a vector of rank the node degree, whose coordinates are the particle colors;
  \item a motion part, which exchanges position states between node \(x\) and its neighbors \cite{Berry-2011qq},
    \begin{equation}
      \label{e:Mop}
      M\ket{xc_ys} = \ket{y c_x s}\,,
    \end{equation}
    for each \(x \in V\) and \(y \in V_x\);
  \item and an interaction part, which includes a color and spin (\(X\)) coupling, and a spin-spin (\(Z\)) coupling; the \(X\) operator exchanges the values of local node color (for the values \(c = 0,1\)) and spin:
    \begin{equation}
      \label{e:Xop}
      X \ket{\ldots c_x \ldots s_x \ldots} = \ket{\ldots s_x \ldots c_x \ldots}
    \end{equation}
  for \(x \in V\), and spin interaction is of the Ising type \cite{Briegel-2001fk},
    \begin{equation}
      \label{e:Zop}
      Z \ket{\ldots s_x \ldots s_y \ldots} = (-1)^{s_xs_y}\ket{\ldots s_x \ldots s_y \ldots}
    \end{equation}
    for \((x,y) \in E\), modifies the phase of two neighboring down spins.
\end{itemize}
Note that the joint action of the \(X\) and \(Z\) operators, creates some frustration in the spin arrangements: the coin-spin interaction flips the local spin, while the Ising interaction, which commutes with \(z\) Pauli matrix, entangles \(x\)-aligned spins; together they act to increase neighboring spins entanglement by favoring equal superposition of up and down orientations. With these definitions the unitary one time step operator is (\(\hbar = 1\)),
\begin{equation}
  \label{e:Uop}
  U \ket{\psi(t + 1)}= Z X M C \ket{\psi(t)}\,,
\end{equation}
where \(t\) is the step number and \(\ket{\psi(t)}\) the corresponding state. The initial state
\begin{equation}
  \label{e:psi0}
  \ket{\psi(0)} = \frac{1}{\sqrt{|V|}} \sum_{x \in V} \ket{x00}
\end{equation}
is chosen to be an equal superposition over the nodes in \(V\), with color \(c=0\), and all spins up \(s=0\).

As already observed in \cite{Verga-2019} the quantum walk evolves towards a thermal state, well described by the microcanonical ensemble. To illustrate the behavior of the thermal state we computed the probability distribution over the nodes of the particle position, the evolution of the mean spin per node and of the entanglement entropy between the tree parts of the Hilbert space \(\{x,c,s\}\) (see Fig.~\ref{f:TH}).

The particle position probability at node \(x\) and time step \(t\) is defined by,
\begin{equation}
  \label{e:pxt}
  p(x,t) = \Tr_{\bar{x}} \rho(t)\,,
\end{equation}
where \(\rho(t) = \ket{\psi(t)}\bra{\psi(t)}\) is the density matrix, and the trace is over the complementary set \(\bar{x}\) of node \(x\) (the other nodes, the color and spin quantum numbers); we note that the numerical computed probabilities match the degree distribution. Indeed, the microcanonical distribution predicts equal probabilities of each state, leading to a homogeneous distribution of the particle positions over the graph according to its connectivity \(p(x) = d_x/\sum_x d_x\). We compare in Fig.~\ref{f:TH} (left), the microcanonical distribution with the numerically computed particle position in the stationary state, reached after an initial transitory, and found a good agreement. 

The mean total spin in the \(z\) direction,
\begin{equation}
  \label{e:sz}
  \bar{s}_z(t) = \Tr_{\bar{s}} \sigma_z \rho(t)\,
\end{equation}
(\(\sigma_z\) is the \(z\) Pauli matrix) represented in Fig.~\ref{f:TH}, evolves from the initial value \(\bar{s}_z(0)=1\) to a paramagnetic \(\bar{s}_z = 0\) state, showing a smooth relaxation.

The entanglement entropy is given by the von Neumann formula,
\begin{equation}
  \label{e:ent}
  S_l(t) = - \Tr \rho_l(t) \log \rho_l(t),\; l = \{x,c,s\}
\end{equation}
where \(\rho_l= \Tr_{\bar{l}} \rho\) is the partial trace of the total density matrix and \(\bar{l}\) is the set \(\{c,s\}\) for the position, the set \(\{x,s\}\) for the color, and \(\{x,c\}\) for the spin entropies, respectively. The base two logarithm is denoted \(\log\). An example of the behavior of the position (\(x\)), color (\(c\)), and spin (\(s\)) entropies is given in Fig.~\ref{f:TH}. It is interesting to note that both, particle distribution and entanglement entropy have a short relaxation time, essentially fixed by the size of the graph (of the order of the ballistic time, \(t\sim |V|\)). In contrast, the spin relaxation is much slower; for the ER graph of Fig.~\ref{f:g}, it take about \num{200} steps.

The operator \(U\) associated with a graph, in particular random graphs, has a set of chaotic eigenvectors and eigenvalues, well described by the unitary Gaussian ensemble (Ref.~\cite{Verga-2019}), which in addition to support the eigenstate thermal hypothesis of quantum isolated systems \cite{Deutsch-1991vn,Srednicki-1994ys}, demonstrates the ability of the quantum walk dynamics to create highly entangled states. In the next section we investigate the structure of the thermal state relating it with the topology of the graph through the entanglement entropy.

\section{Graph cycles and entanglement entropy}

\begin{figure}
  \centering
  \includegraphics[width=0.49\textwidth]{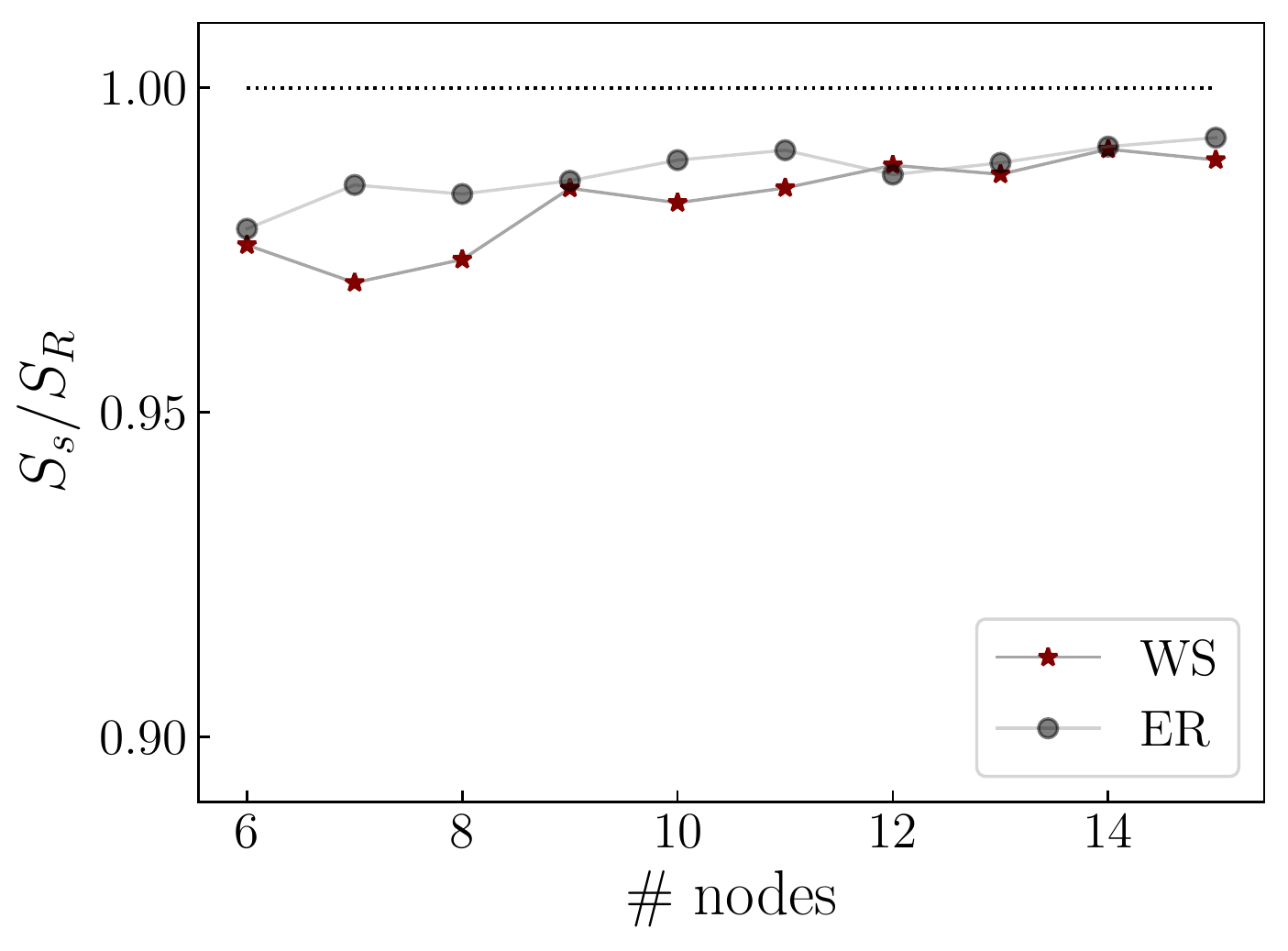}
  \caption{Spin entanglement entropy $S_s$, normalized to the Page entropy for a pure random state, as a function of the node number of graphs WS and ER.
  \label{f:page}}
\end{figure}

The entanglement entropy can be viewed as a measure of the correlations between the degrees of freedom of a many-body complex quantum state, beyond the classical ones. However, the entanglement entropy can in addition, unveil topological properties of the quantum state. Indeed, the entanglement entropy is used to characterize topological phases in quantum liquids \cite{Kitaev-2006uq,Levin-2006fk}. For instance, in the case of the toric code, the entanglement entropy between the interior of a connected region and its exterior, has two contributions: one proportional to the perimeter (area law, in two dimensions) of the region, and the other, nonlocal, which depends on the topology of the region \cite{Hamma-2005} (for example, its connectivity). The opposite situation would be the one of pure random states, or typical states in the context of eigenvectors thermalization. Page \cite{Page-1993nr} computed the entanglement entropy in a pure random state, and showed that, for a subsystem \(A\), it is proportional to the logarithm of its dimension \(D_A\) and a correcting term,
\begin{equation}
\label{e:page}
S_R(A) = \log D_A - \frac{D_A^2}{2D\ln(2)}\,.
\end{equation}
(we use throughout base 2 logarithms) where \(D=|\mathcal{H}|\) is the dimension of the total system. This formula, which applies to thermal states \cite{Zhang-2015}, gives a good estimation of the entanglement in chaotic systems \cite{Mejia-Monasterio-2005}. It shows that the entropy of a subsystem in a typical state is proportional to its volume (its number of degrees of freedom), and that, for such states, it is maximal.

We have shown that the interacting quantum walk reach an equilibrium state well described by a thermal ensemble (previous section and Ref.~\cite{Verga-2019}). The rules defining the system's evolution \eqref{e:Cop}-\eqref{e:Zop}, can be thought as an algorithm to build up, starting with a simple product state, an entangled state. The rules are local, only neighboring spins interact, and the particle walk is also between neighboring nodes. An interesting question arises about the entanglement structure of the thermal state. In particular, one may ask weather the entangled state can be compared with the graph states \cite{Hein-2006eu}, or with the ground state of the spin liquid topological phases \cite{Zeng-2019}. Indeed, as in the construction of a graph state \cite{Briegel-2001fk}, the joint action of \(C\) and \(Z\) tends to maximize the entanglement of interacting spins. For instance, in the Kitaev toric code \cite{Kitaev-2003fk}, the ground state is an equal superposition of closed strings in the square lattice of, say, spins down in a background of spins up. Even if this comparison may seem unnatural, spin liquids follows an area law for entanglement at variance to the volume law expected for our thermal state, both systems possess random states and an analogy could be interesting. In our system, the particle explores the graph creating correlations by interacting with the local spins, favoring states of maximal entanglement, as we discussed in \S~\ref{s:model}.

We can test the randomness of the quantum state reached after \(t = 400\) iterations \(U\), by comparing the numerical values of the spin entanglement entropy with the Page entropy; their ratio should equal one for a pure random state of the particle-spin system. In Fig.~\ref{f:page} we show the result; it confirms that for the set of random graphs studied, the system reach a near random, thermal state (see also Fig.~\ref{f:TH}). The question arises, and this is our central point, whether this random state has, as for instance a spin liquid ground state, a specific hidden structure associated with the underlying graph topology. 

Assuming that the thermal state generated by \(U\) is dominated by a superposition of closed spin strings maximally entangled with the walker, we can guess an explicit expression of the entanglement entropy, and compare it with the exactly computed one. We need before to define the cycle structure of a connected simple graph.

\begin{figure*}
  \centering
  \includegraphics[width=0.33\textwidth]{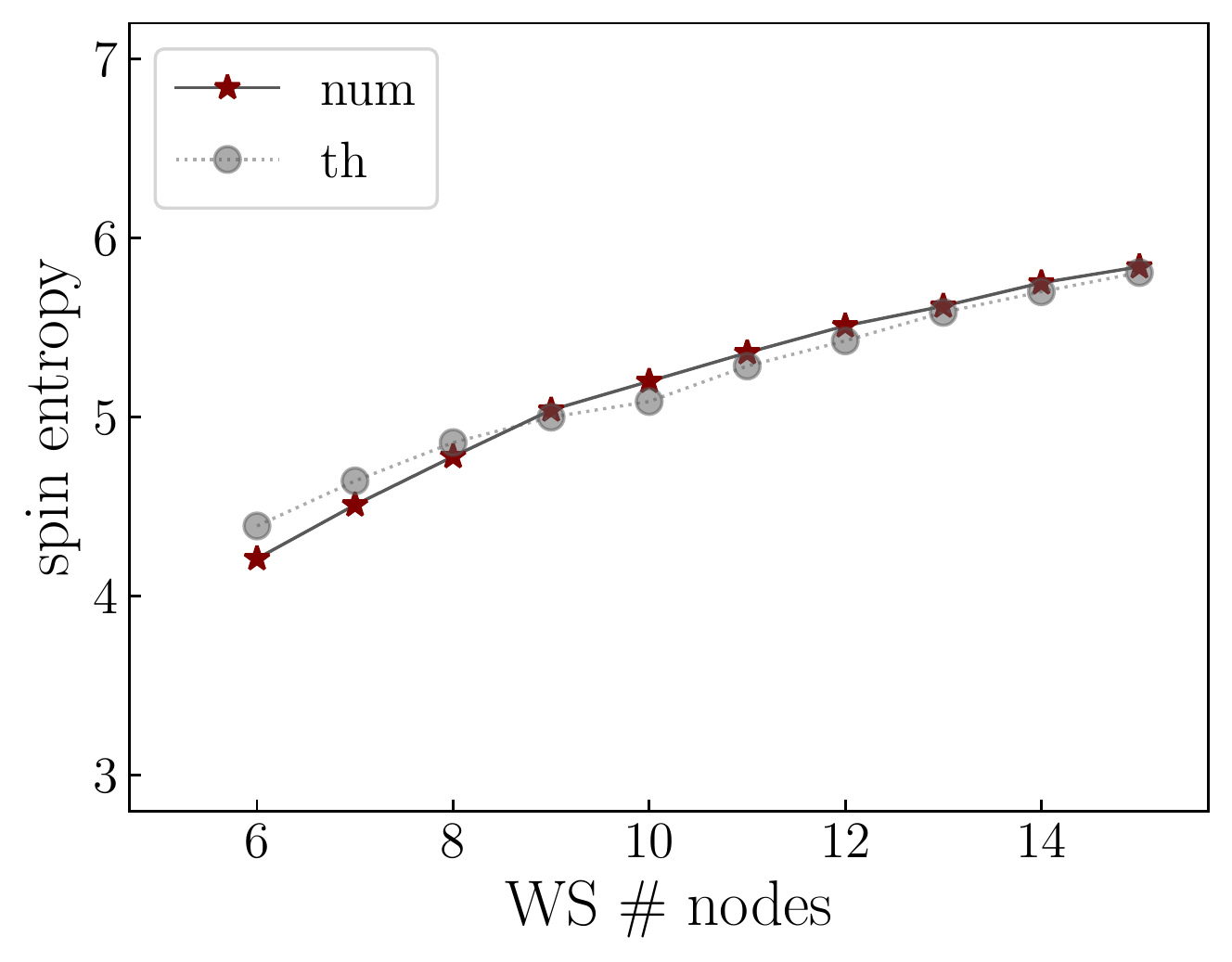}%
  \includegraphics[width=0.33\textwidth]{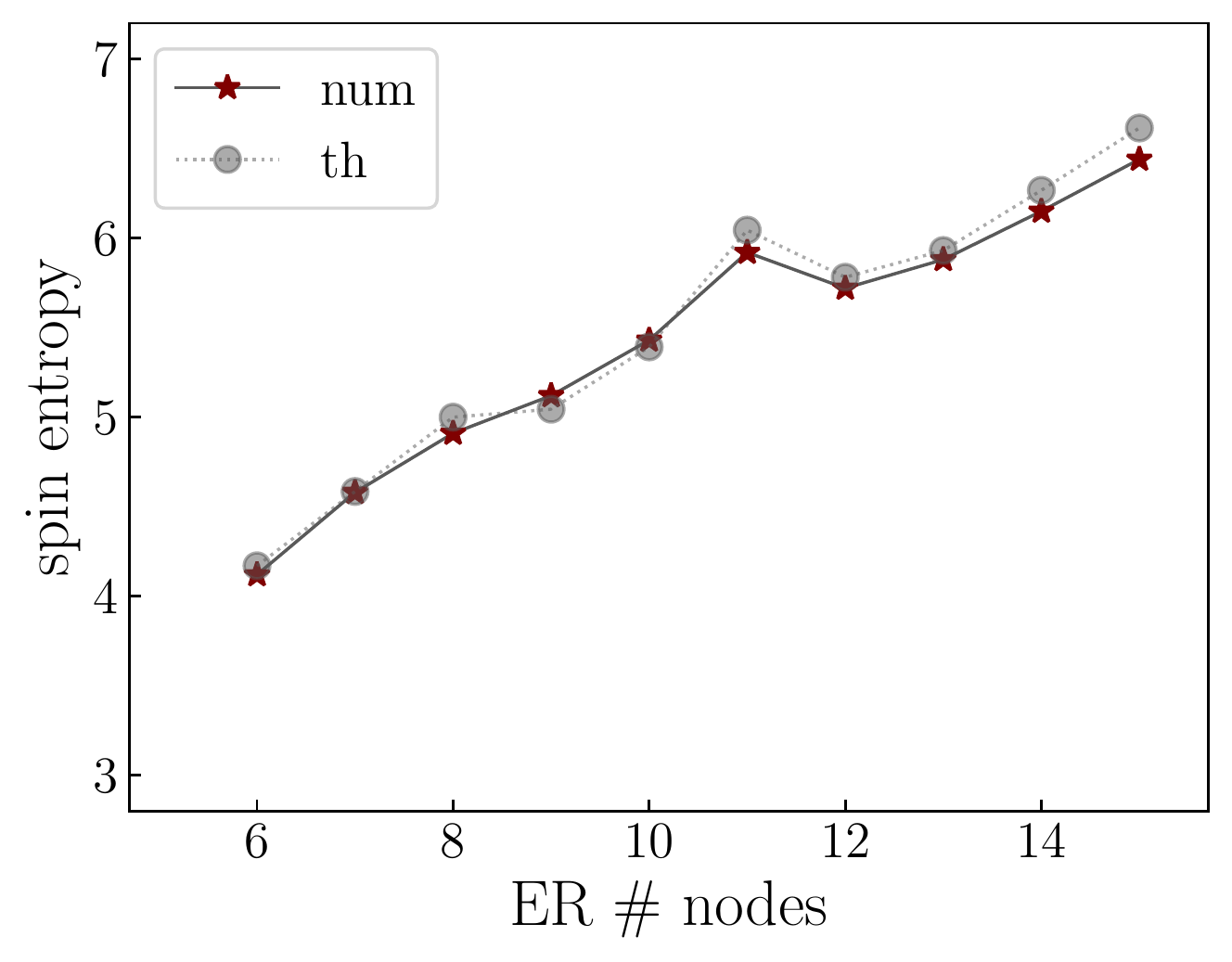}%
  \includegraphics[width=0.33\textwidth]{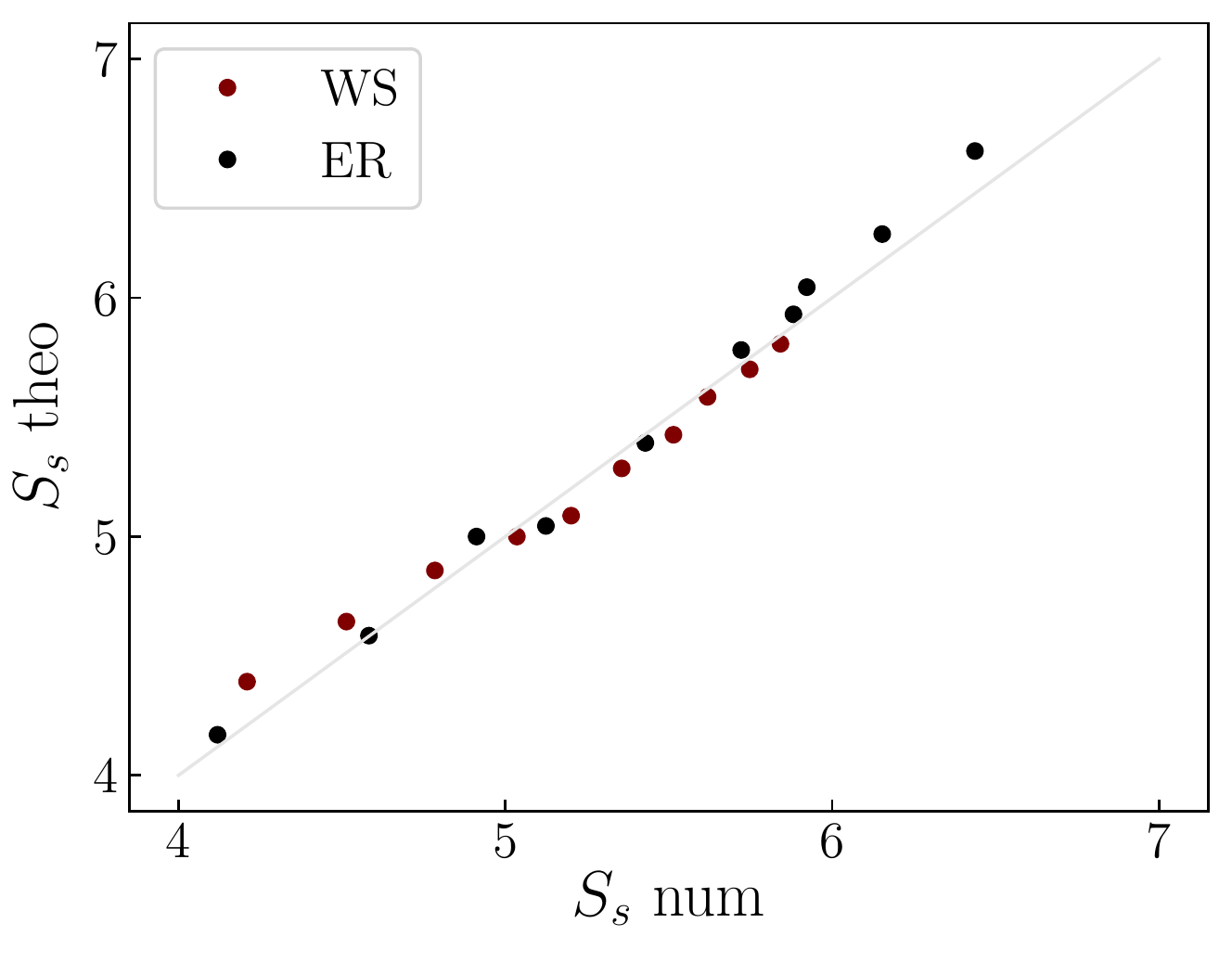}
  \caption{Spin entanglement entropy for random graphs with 6 to 15 nodes; (left) WS, (center) ER. Numerical results are compared with formula \protect\eqref{e:Ss}; (left) Comparison of the numerical (abscissa) and theoretical (ordinate) spin entanglement entropies for the WS and ER graphs from 6 to 15 nodes.
  \label{f:ntR}}
\end{figure*}

A cycle \(b\) is a closed path in \(G\); it is a subset of the graph edge set \(E\). The set \(B_C(G)\) of all cycles is the cycle space. To each cycle \(b \in B_C(G)\) we can associate a vector with \(|E|\) components, each taking the values in the set \(\{0,1\}\), where the value \(1\) stands for an edge in \(b\), and \(0\) otherwise. The cycle space \(B_C(G)\) equipped with the ring sum forms a vector space over the finite field \(\mathrm{GF}(2)\), of dimension \cite{Gross-2005},
\begin{equation}
  \label{e:dimB}
  |B| = |E| - |V| + 1
\end{equation}
(for a connected graph). The composition rule of the vector space, the ring sum \(b_1 \oplus b_2\), corresponds to the symmetric difference between the edges subsets \((b_1 \cup b_2) \setminus (b_1 \cap b_2)\). A set of cycles that cannot be written one another as a ring sum combination, is an independent set. A set \(B \subset B_C\) of independent cycles of dimension \(|B|\), given by \eqref{e:dimB}, is a cycle basis \(B\):
\begin{equation}
  \label{e:setB}
  B = \{b_n \in B_C,\; n=1,\ldots,|B|\}\,,
\end{equation}
of \(B_C\). Therefore, every element \(b\in B_C\) can be written as a linear combination of the basis cycles:
\[b = \sum_{n = 1}^{|B|} s_n b_n,\quad s_n = 0, 1\,.\]
We call the minimum cycle basis of \(B_C\), the basis of shortest length \cite{Kavitha-2009}, 
\begin{equation}
  \label{e:Bmin}
  B_G = \left\{ b_n \in B \,|\, \min \len(B) \right\}
\end{equation}
where,
\[\len(B) = \sum_{n=1}^{|B|} \len(b_n) \]
with the length of a cycle \(b\) defined as the total number of its nodes:
\[\len(b) = |b|\,.\]
The reason we define the weight of a basis as its number of nodes is because the physical degrees of freedom associated with the particle and spins, reside on the nodes.

The fact that the particle-spin entanglement could be dominated by closed strings naturally should establish a relation with the graph cycle basis, which characterizes the graph organization by an independent set of elementary structures. Indeed, it is natural to think that each basis cycle corresponds to a distinct configuration of spins whose overlap with the particle state is proportional to the vertices in the cycle. We therefore propose a formula of the spin entanglement entropy in terms of the cycle basis with minimum length:
\begin{equation}
  \label{e:Ss}
  S_s \approx  S_C = \log\left[ \sum_{n = 1}^{|B|} \len(b_n)\right]\,,
\end{equation}
where \(b_n \in B_G\), applicable when the system is in its thermal state.

We test now conjecture \eqref{e:Ss} on the series of random graphs introduced before (WS and ER). For each graph, with vertex number \(|V| = 6,\ldots,15\), we measure the entanglement entropy and compare it with formula \eqref{e:Ss} (Fig.~\ref{f:ntR}). The main difference between the two types of graphs, is that WS has fixed mean degree \(\bar{d} = 4\), while it varies for ER graphs \(\bar{d}=\{3.7,4,4.3,4.2,4.6,5.8,4.7,4.8,5.3,6\}\); as a consequence WS has, for the same number of vertices, a smaller spanning basis than ER. The richer cycle structure of the ER graphs reflects in the faster growth of the entanglement with the number of nodes than for the WS graphs (compare Fig.~\ref{f:ntR} left and middle). This shows that the spin entanglement is not directly proportional to simply the number of nodes but instead, with the cycle basis rank and its length (note the slope inversion at \(|V|=11\) in Fig.~\ref{f:ntR}, which corresponds to a locally larger mean degree and basis size at \(|V| = 11\) node graph, larger than the one for the \(|V| = 12\) graph). We collected in Fig.~\ref{f:ntR} (right) all numerical and theoretical values of \(S_s\) showing that the numerical computed spin entanglement entropy agrees within \(2\%\) error with \eqref{e:Ss}.

Therefore, we find that for a variety of graphs, with a large range of Hilbert space dimensions and topologies, the conjecture \eqref{e:Ss} suitably describes the quantitative behavior of the particle-spin entanglement in the thermal state.

\section{Conclusions}

Using a model of an interacting quantum system introduced in \cite{Verga-2019}, we investigated the entanglement structure of the thermal state. The model do not contain dimensional parameters: it is essentially defined by the graph. The physical properties of the particle-spin interacting system reflect the topological structure of the graph. In this sense, we can say that the thermal state generated by \(U\), at variance to a pure uncorrelated random state, is chaotic and should possess some internal structure. Assuming that the interaction between the walker and the local spins favors a superposition of states associated with the topology of the graph, in particular its cycle vector space structure, we conjectured a formula for the entanglement entropy. We found that the hypothesis of an entangled state as a superposition of closed spin strings following the graph cycles, leads to a formula Eq.~\eqref{e:Ss}, of the particle-spin entanglement entropy, which is well verified by numerical computations.

The specific correlations found in the thermal state, can be traced back to the way the particle walks on the graph, interacting through its color with the local spins, following essentially one dimensional paths (only the values of \(c=0,1\) intervene in the particle-spin interaction). However, we would think that the universality of the statistical properties of the evolution \(U\) ensures the validity of our findings about the existence of a characteristic structure behind the randomness of the thermal state, to other systems satisfying the eigenvalue thermalization and state typicality.

\begin{acknowledgments}
  We greatly appreciated discussions with Laurent Raymond and Pablo Arrighi. AV thanks the kind invitation of the Departamento de Física, Universidad de Santiago, where part of this work was realized. We acknowledge support by Fondecyt Iniciación No.\ 11171122 and CEDENNA, Universidad de Santiago.
\end{acknowledgments}

%
%\bibliography{total}
%
%merlin.mbs apsrev4-1.bst 2010-07-25 4.21a (PWD, AO, DPC) hacked
%Control: key (0)
%Control: author (8) initials jnrlst
%Control: editor formatted (1) identically to author
%Control: production of article title (-1) disabled
%Control: page (0) single
%Control: year (1) truncated
%Control: production of eprint (0) enabled
%
%
%
\end{document}